\documentclass[aps,prb,twocolumn,groupedaddress]{revtex4-1}
\usepackage[dvipdfmx]{graphicx}
\usepackage{amsmath,amssymb}
\usepackage{mathrsfs}
\usepackage{bm}
\usepackage{braket}
\begin{document}


\title{Long time asymptotic state of periodically driven open quantum systems}
\author{Koudai Iwahori}
\email[]{iwahori@scphys.kyoto-u.ac.jp}
\author{Norio Kawakami}
\affiliation{Department of Physics, Kyoto University, Kyoto 606-8502, Japan}
\date{\today}

\begin{abstract}
We investigate a long time asymptotic state of periodically driven open quantum systems analytically.
The model we consider in this paper is a free fermionic system coupled to an energy and particle reservoir.
We clarify  some generic properties of the system which are independent of the details of the reservoir in the high frequency regime of the external driving.  When the frequency of the external driving is much larger than the energy cutoff of the system-reservoir coupling, the low-energy properties of the system are equivalent to those of the Gibbs distribution of a Floquet effective Hamiltonian. Furthermore, we investigate the effect of finite dissipation on the system, and elucidate that we cannot suppress excitations by merely increasing the system-reservoir coupling because the excitations are created through the reservoir by the external driving when the frequency is smaller than the energy cutoff of the system-reservoir coupling. 
\end{abstract}


\maketitle

\section{Introduction}
Fast oscillating external drivings have become an important tool to manipulate quantum phases of matter.
By irradiating intense laser lights to solid state systems \cite{Oka2009,Lindner2011,Kitagawa2011} or shaking optical lattices in ultracold atomic systems \cite{Linger2007,Struck2011,Aidelsburger2013a,Aidelsburger2013b,Miyake2013,Aidelsburger2015} 
we can realize novel properties of matter which the system does not have before applying external drivings, and these phenomena have been observed experimentally.

Among various intriguing topics, long time dynamics of periodically driven quantum systems has attracted particular attention in condensed matter physics.
Theoretical treatments of periodically driven systems rely on the Floquet theory which ensures that the long time dynamics can be described by the static effective Hamiltonian \cite{Shirley1965,Sambe1973}.
However, this fact does not necessarily lead to the equivalence between the properties of the long time asymptotic state and the thermodynamic properties of the effective Hamiltonian.
Recently, the essential feature of periodically driven {\it isolated systems} has been revealed \cite{Lazarides2014a,DAlessio2013,Lazarides2014b,Ponte2015,Abanin2015a,Abanin2015b,Abanin2015c,Mori2016,Mori2015a,Bukov2015,Kuwahara2016,Das2010,Bhattacharyya2012,Hegde2014}; it has been shown that while integrable systems can be described by the time-periodic generalized Gibbs state \cite{Lazarides2014a}, nonintegrable systems heat up and show infinite temperature behaviors after a long time \cite{DAlessio2013,Lazarides2014b,Ponte2015}.
Further studies have also shown that the energy absorption rate is exponentially small in the high frequency regime of the external driving \cite{Abanin2015a,Abanin2015b,Abanin2015c,Mori2016}. 

In spite of these intensive studies, there still remain fundamental problems in periodically driven quantum systems.
In particular, {\it open quantum systems} in which the system under consideration is coupled to a thermal bath have been investigated extensively \cite{Kohler1997,Breuer2000,Kohn2001,Hone2009,Ketzmerick2010,Iadecola2013,Langemeyer2014,Iadecola2015a,Iadecola2015b,Shirai2015a,Liu2015,Seetharam2015,Dehghani2014}.
However, general properties common to periodically driven open systems have not been understood well.
Without external driving, the detailed balance condition ensures that the asymptotic state of the system can be described by the Gibbs state.
On the other hand, when the external time periodic driving is applied, the detailed balance condition is not fulfilled, and therefore the asymptotic state cannot be described by the Gibbs distribution of the effective Hamiltonian (Floquet Gibbs state) and depends on the details of the reservoir in general.
It has been shown that if certain special conditions are satisfied, the detailed balance condition is fulfilled and therefore the long time asymptotic state of periodically driven open quantum systems can be described by the Floquet Gibbs state \cite{Shirai2015a,Liu2015,Iadecola2015a}.
However, these conditions require fine tuning of the system-reservoir coupling, and cannot be applied for generic cases.
In addition, these studies have treated the weak system-reservoir coupling limit, and thus the effect of finite dissipation has not been taken into account. 
Recent study by Shirai {\it et al.} \cite{Shirai2016} has shown numerically that the conditions obtained in the paper \cite{Shirai2015a} can be relaxed, but their analysis still requires another fine tuning for external field.
Therefore it is desirable to understand general properties expected for periodically driven open quantum systems, which are not specific to the special models.

In this paper, we explore some generic properties of the periodically driven open free fermionic systems, which are independent of the details of the reservoir, and also clarify the effect of finite dissipation.
Our results reveal that 
low-energy properties of long time asymptotic states are generally described by those of the Floquet Gibbs state if the frequency of the external driving is much larger than the energy cutoff of the system-reservoir coupling.
We also find that excitations cannot be suppressed by simply rendering the system-reservoir coupling stronger.
These facts demonstrate that in order to obtain the equilibrium properties of a Floquet effective Hamiltonian in open quantum systems, we need high frequency external drivings whose frequency exceeds the energy cutoff of the system-reservoir coupling.

In the next section, we show the model considered in this paper and derive the equation of motion of the system.
We investigate the long time asymptotic state of the system in the weak system-reservoir coupling limit in Sec. \ref{Sec_WeakCouplingLimit},
then we discuss the effect of finite system-reservoir coupling in Sec. \ref{Sec_FiniteCoupling}.
A summary of our results is presented in Sec. \ref{Sec_Summary}.

\section{Setup}
Let us start with a setup of our system.
The total Hamiltonian is given by 
\begin{equation}
	H_{\text{tot}}(t)=H_{S}(t)+H_{R}+H_{I},
\end{equation}
where $H_{S}(t)$ and $H_{R}$ are the Hamiltonian of the system and the reservoir, and $H_{I}$ represents the interaction between them.
The Hamiltonian of the system is bilinear $H_{S}(t)=\sum_{\alpha,\beta}H_{\alpha\beta}(t)a^{\dagger}_{\alpha}a_{\beta}$, where $a_{\alpha}$ is an annihilation operator of a fermion in the system.
The time dependent matrix $H_{\alpha\beta}(t)$ has a period of $\tau$: $H_{\alpha\beta}(t+\tau)=H_{\alpha\beta}(t)$ and the degree of freedom of the system is finite.
We take the reservoir as a fermionic bath $H_{R}=\sum_{k}\epsilon_{k}A^{\dagger}_{k}A_{k}$, and consider the continuum limit $\sum_{k}\frac{2\pi}{L}F(\epsilon_{k})=\int_{-\infty}^{\infty}d\epsilon D(\epsilon)F(\epsilon)$, where $D(\epsilon)$ is the density of states of the reservoir and $F(\epsilon)$ is an arbitrary function.
The initial state of the reservoir is the equilibrium state with the temperature $T$ and the chemical potential $\mu$, that is, $\braket{A^{\dagger}_{k}A_{k}}=f_{T}(\epsilon_{k}-\mu)$, where $f_{T}(\omega)=1/(e^{\omega/T}+1)$ is the Fermi distribution function.
Without loss of generality, the chemical potential $\mu$ can be taken as zero.
The system-reservoir coupling, which controls the exchange of particles as well as energy, is assumed to be bilinear $H_{I}=\sum_{k,\alpha}(\lambda_{k,\alpha}A^{\dagger}_{k}a_{\alpha}+h.c.)/\sqrt{L}$.

The Heisenberg equation of the total system reads
\begin{align}
	&i\frac{d}{dt}A_{k}(t)=\epsilon_{k}A_{k}(t)+\sum_{\alpha}\frac{\lambda_{k,\alpha}}{\sqrt{L}}a_{\alpha}(t)\label{Eq_ReservoirEOM}\\
	&i\frac{d}{dt}a_{\alpha}(t)=\sum_{\beta}H_{\alpha\beta}(t)a_{\beta}(t)+\sum_{k}\frac{\lambda_{k,\alpha}^{\ast}}{\sqrt{L}}A_{k}(t)\label{Eq_SystemEOM}.
\end{align}
Integrating eq. (\ref{Eq_ReservoirEOM}) and substituting into eq. (\ref{Eq_SystemEOM}), we have a closed form of the equation of motion for the system:
\begin{equation}
	\begin{aligned}
		i\frac{d}{dt}a_{\alpha}(t)=&\sum_{\beta}H_{\alpha\beta}(t)a_{\beta}(t)\\&-i\sum_{\beta}\int_{0}^{t}dt^{\prime}\Gamma_{\alpha\beta}(t^{\prime})a_{\beta}(t-t^{\prime})+i\xi_{\alpha}(t),\label{Eq_SystemClosedEOM}
	\end{aligned}
\end{equation}
where $\Gamma_{\alpha\beta}(t)=\sum_{k}\frac{\lambda_{k,\alpha}^{\ast}\lambda_{k,\beta}}{L}e^{-i\epsilon_{k}t}=\int\tilde{\Gamma}_{\alpha\beta}(\omega)e^{-i\omega t}d\omega$ and $\xi_{\alpha}(t)=-i\sum_{k}\frac{\lambda_{k,\alpha}^{\ast}}{\sqrt{L}}e^{-i\epsilon_{k}t}A_{k}$.
In addition to the unitary time evolution by the Hamiltonian $H_{\alpha\beta}(t)$, eq. (\ref{Eq_SystemClosedEOM}) contains the dissipation term $\Gamma_{\alpha\beta}(t)$ and the noise term $\xi_{\alpha}(t)$.

\section{Results for weak system-reservoir coupling limit}\label{Sec_WeakCouplingLimit}
\subsection{Derivation of long time asymptotic states in the weak system-reservoir coupling limit}
From now on, we write the time periodic matrix $H_{\alpha\beta}(t)$ as $[H(t)]_{\alpha\beta}$.
We here make use of Floquet theorem to treat the time dependence in eq. (\ref{Eq_SystemClosedEOM}). Floquet theorem ensures the existence of a time periodic unitary transformation $W_{\alpha\beta}(t)=W_{\alpha\beta}(t+\tau)$ which eliminates the periodic time dependence of the Hamiltonian $H_{\alpha\beta}(t)$: 
\begin{equation}
	H^{\text{eff}}=W^{\dagger}(t)H(t)W(t)-W^{\dagger}(t)i\frac{d}{dt}W(t).
\end{equation}
Diagonalize the effective Hamiltonian $H^{\text{eff}}$ by a unitary transformation $V$ and define an annihilation operator of the Floquet state, $c_{\alpha}(t)=\sum_{\beta}U^{\dagger}_{\alpha\beta}(t)a_{\beta}(t)$, where $U(t)=W(t)V$.
Then the equation of motion of the Floquet state follows from eq. (\ref{Eq_SystemClosedEOM}),
\begin{equation}
	\begin{aligned}
		&i\frac{d}{dt}c_{\alpha}(t)=\epsilon_{\alpha}c_{\alpha}(t)+i\eta_{\alpha}(t)\\
									&-i\sum_{\beta}\sum_{n,m\in\mathbb{Z}}e^{-i(n-m)\Omega t}\int_{0}^{t}dt^{\prime}\Gamma_{\alpha\beta}^{nm}(t^{\prime})e^{-im\Omega t^{\prime}}c_{\beta}(t-t^{\prime}),\label{Eq_FloquetEOM}
	\end{aligned}
\end{equation}
where $\eta_{\alpha}(t)=\sum_{\beta}U^{\dagger}_{\alpha\beta}(t)\xi_{\beta}(t)$, $\Omega=2\pi/\tau$ and $\Gamma_{\alpha\beta}^{nm}(t)=[U^{(n)\dagger}\Gamma(t)U^{(m)}]_{\alpha\beta}$.
$\epsilon_{\alpha}$ is an eigenvalue of the effective Hamiltonian $H^{\text{eff}}$ and $U^{(n)}=\int_{0}^{\tau}U(t)e^{in\Omega{t}}dt/\tau$.

Let us now consider eq. (\ref{Eq_FloquetEOM}) in the weak coupling limit of the system-reservoir coupling (van Hove limit \cite{VanHove1955}), as done in many of the previous papers \cite{Shirai2015a,Liu2015,Iadecola2015a}.
In this limit, the time evolution of the system of interest follows the Markovian quantum master equation.
Here, we assume the conditions of nondegeneracy and nonresonance ($\epsilon_{\alpha}-\epsilon_{\beta}+n\Omega=0 \Leftrightarrow \alpha=\beta,\ n=0$).
Then define the dissipation rate by $\gamma :=\int_{0}^{\infty} \Gamma(t)dt$ and take the limit of $\lim_{\gamma t\rightarrow \infty}\lim_{\gamma/(\epsilon_{\alpha}-\epsilon_{\beta}+n\Omega)\rightarrow 0}$, 
that is, take the weak system-reservoir coupling limit $\gamma/(\epsilon_{\alpha}-\epsilon_{\beta}+n\Omega)\rightarrow 0$ ($\alpha\neq\beta$)  first, then take the long time limit $\gamma t\rightarrow \infty$.
In this limit, only the diagonal part of the dissipation term $\Gamma_{\alpha\beta}^{nm}(t)$ ($\alpha=\beta$, $n=m$) contributes to the results and eq. (\ref{Eq_FloquetEOM}) can be solved as
\begin{equation}
	c_{\alpha}(t)=\sum_{k,\beta}\sum_{n\in\mathbb{Z}}\frac{1}{\sqrt{L}}\frac{U^{(n)\dagger}_{\alpha\beta}(\lambda_{k,\beta})^{\ast}e^{-i(\epsilon_{k}-n\Omega)t}}{\epsilon_{k}-\epsilon_{\alpha}^{(n)}+i\Gamma_{\alpha}^{\prime}(n\Omega-\epsilon_{k})}A_{k}, \label{Eq_SolutionOfEOM}
\end{equation}
where $\Gamma_{\alpha}^{\prime}(\omega)=\sum_{n\in\mathbb{Z}}\int_{0}^{\infty}dt\Gamma_{\alpha\alpha}^{nn}(t)e^{i(\omega-n\Omega)t}$ and $\epsilon_{\alpha}^{(n)}=\epsilon_{\alpha}+n\Omega$.
Consequently, we end up with the simple expression for the occupation number of the Floquet state,
\begin{equation}
	\braket{c^{\dagger}_{\alpha}(t)c_{\alpha}(t)}=\frac{\sum_{n\in\mathbb{Z}}\Gamma_{\alpha}^{n}f_{T}(\epsilon_{\alpha}+n\Omega)}{\sum_{n\in\mathbb{Z}}\Gamma_{\alpha}^{n}},\label{Eq_FlouetOccupationNum}
\end{equation}
where $\Gamma^{n}_{\alpha}=[U^{(n)\dagger}\tilde{\Gamma}(\epsilon_{\alpha}^{(n)})U^{(n)}]_{\alpha\alpha}$ (for the details of the calculation, see the Appendix \ref{appendix_a}).
The occupation number of the Floquet state is time independent.
The similar expression can be obtained for the time correlation functions.
We can easily extend this formula to the multiple reservoirs cases, and in the two reservoirs case, our result is reduced to the one obtained by Iadecola and Chamon\cite{Iadecola2015a}.

Obviously, eq. (\ref{Eq_FlouetOccupationNum}) is different from the occupation number obtained for the Floquet Gibbs state in general.
The expression (\ref{Eq_FlouetOccupationNum}) allows us to regard the driven  system as a sum of infinite number of bands \cite{Tien1963} which are shifted by $n\Omega$ because there is a sum of the Fermi distribution functions in the numerator and each Fermi distribution function is shifted by $n\Omega$ in eq. (\ref{Eq_FlouetOccupationNum}).
We call the shifted band as the $n$th sideband.
The contribution from the $n$th sideband to the occupation number is determined by the weight $w_{\alpha}^{n}=\Gamma_{\alpha}^{n}/\sum_{m\in\mathbb{Z}}\Gamma_{\alpha}^{m}$.
This expression is invariant under the gauge transformation $G(t)=\exp[iN\Omega t], N_{\alpha\beta}=n_{\alpha}\delta_{\alpha\beta}\ (n_{\alpha}\in\mathbb{Z})$ which transforms $\epsilon_{\alpha}$ and $\Gamma^{n}_{\alpha}$ into $\epsilon_{\alpha}+n_{\alpha}\Omega$ and $\Gamma^{n+n_{\alpha}}_{\alpha}$.

We here examine, in the light of eq. (\ref{Eq_FlouetOccupationNum}), what is meant by the previously proposed special conditions for the emergence of the Floquet Gibbs state \cite{Shirai2015a,Liu2015,Iadecola2015a}. If those conditions are applied for our results, only the $0$th band contributes to the occupation number, that is, $w_{\alpha}^{n}=0$ for $n\neq 0$, and
thereby the system is indeed described by the Floquet Gibbs state rigorously $\braket{c^{\dagger}_{\alpha}(t)c_{\alpha}(t)}=f_{T}(\epsilon_{\alpha})$.
This means that the $0$th band describes the Floquet Gibbs state while the other sidebands describe the deviation from the Floquet Gibbs state.
When we can apply the Magnus expansion \cite{Magnus1954,Blanes2009} to obtain the quasi-energy spectrum, $U^{(n)}=O((A/\Omega)^{n})$, the deviation from the Floquet Gibbs state is $O((A/\Omega)^2)$, where $A$ is the amplitude of the external driving.
In our expression the conditions for the emergence of the Floquet Gibbs state can be written as $[U^{(n)\dagger}\tilde{\Gamma}(\omega)U^{(n)}]_{\alpha\alpha}=0$ for $n\neq 0$.

\subsection{Generic properties of the system in the high frequency regime of the external field}
We are now ready to extract general properties of the long time asymptotic state from eq. (\ref{Eq_FlouetOccupationNum}) in the high frequency regime.
We show below that the low-energy properties are equivalent to those of the Floquet Gibbs state if the frequency of the external driving is much larger than the energy cutoff of the system-reservoir coupling $\omega_{c}$.
The low-energy means the energy shell of $|\epsilon| < \omega_{c}$ because the system is coupled to a reservoir whose chemical potential is taken as zero. In other words, the chemical potential plays a role as the origin of the energy.
Then, we take the quasi-energy spectrum as $-\Omega/2 \leq \epsilon_{\alpha} < \Omega/2$, and these definitions, we obtain the result mentioned above.

We introduce the energy cutoff of the system-reservoir coupling $\omega_{c}$ (or equivalently energy cutoff of the reservoir) such that $|\tilde{\Gamma}(E)|/|\tilde{\Gamma}(\epsilon)|=O(\omega_{c}/E)$, where $|\epsilon| < \omega_{c}$ and $|E|\gg\omega_{c}$.
This means that the coupling between the system and the states with energy $\omega$ in the reservoir is cut off by $\omega_{c}$ smoothly because $\tilde{\Gamma}(\omega)$ determines the coupling strength between the system and the states with energy $\omega$ in the reservoir.
We consider the case where the frequency of the external driving is much larger than the energy cutoff of the system-reservoir coupling $\Omega \gg \omega_{c}$.
We then classify the Floquet states into two regions;
(i) states with $|\epsilon_{\alpha}|<\omega_{c}$,
(ii) the other states, i.e. states with $|\epsilon_{\alpha}|\geq \omega_{c}$.
In this case, if $U^{(0)} \gtrsim U^{(n\neq 0)}$ is satisfied, the occupation number of the Floquet states in the region (i) is written as 
\begin{equation}
	\braket{c^{\dagger}_{\alpha}(t)c_{\alpha}(t)}=f_{T}(\epsilon_{\alpha})+O(\omega_{c}/\Omega). \label{Eq_ApploxOccupationNum}
\end{equation}
This is because the contribution from the $n$th sideband is written as $\Gamma_{\alpha}^{n}=[U^{(n)\dagger}\Gamma(\epsilon_{\alpha}+n\Omega)U^{(n)}]_{\alpha\alpha}$ and $|\Gamma(\epsilon_{\alpha})|/\Gamma(\epsilon_{\alpha}+n\Omega)|=O(\omega_{c}/\Omega)$.
When we can apply Magnus expansion to obtain the quasi-energy spectrum, $U^{(n)}$ is obtained as $U^{(n)}=O((A/\Omega)^{n_{0}+n})+O((\Delta/\Omega)^{n_{0}+n})$, where $A$ is the amplitude of the external field and $\Delta$ is the band width of the system.
If the energy spectrum of the system without the external field lies in the range from $(N-1/2)\Omega$ to $(N+1/2)\Omega$, we have $n_{0}=N$.
Therefore, if the chemical potential of the reservoir lies in the energy spectrum of the system, $U^{(0)}\gtrsim U^{(n\neq 0)}$ is satisfied and the occupation number of the Floquet states in the region (i) is written as eq. (\ref{Eq_ApploxOccupationNum}).
We can show similar results for the case where we need a time periodic unitary transformation in order to apply Magnus expansion. This case includes the resonant driving cases and the $A \gtrsim \Omega$ cases such as the system for which the dynamical localization occurs. 
In this case, $U(t)$ is written by the product of two time periodic unitary transformations $U(t)=V(t)W(t)$. $V(t)$ transforms the original time periodic Hamiltonian into the form to which the Magnus expansion is applicable.
In the resonant driving cases, $V(t)$ eliminates the resonance and in the dynamical localization case, $V(t)=\exp[-i\int_{0}^{t}H_{\text{ext}}(\tau)d\tau]$. After applying this unitary transformation $V(t)$, we can apply Magnus expansion, and thereby $W^{(n)}=O((\Delta/\Omega)^n)$ and $U^{(n)}=V^{(n)}+O(\Delta/\Omega)$, where $\Delta$ is bandwidth of the system after applying the time-periodic unitary transformation $V(t)$.
Therefore, if $|V^{(0)}|\gtrsim |V^{(n\neq 0)}|$ (this is satisfied in the resonant driving cases and the dynamical localization case) is satisfied, the Floquet states in the region (i) is written by eq. (\ref{Eq_ApploxOccupationNum}).
Above discussion does not depend on the details of the reservoir. However, the occupation number of states in the region (ii) depends on the details of the reservoir in general and we cannot extract the properties which are independent of the details of the reservoir.

From these observations, we come to the conclusion that the occupation number of states in the region (i) which controls the low-energy properties of the system is given by that of the Floquet Gibbs state when the frequency of the external driving is much larger than the energy cutoff of the system-reservoir coupling.
The same thing can be also proven for the time correlation functions.
Therefore, in order to obtain the low-energy properties of the Floquet Gibbs state, we merely need high frequency external drivings so that the frequency exceeds the energy cutoff of the system-reservoir coupling.
Fine tuning of the system-reservoir coupling, which was required for the previous treatments \cite{Shirai2015a,Liu2015,Iadecola2015a}, is not necessary in this case.
Our results are consistent with the numerical results by Shirai {\it et al.} \cite{Shirai2016} and do not require the special condition which Shirai {\it et al.} \cite{Shirai2016} assumed.

\subsection{Example}
\begin{figure}[t]
	\centering
	\includegraphics[width=80mm]{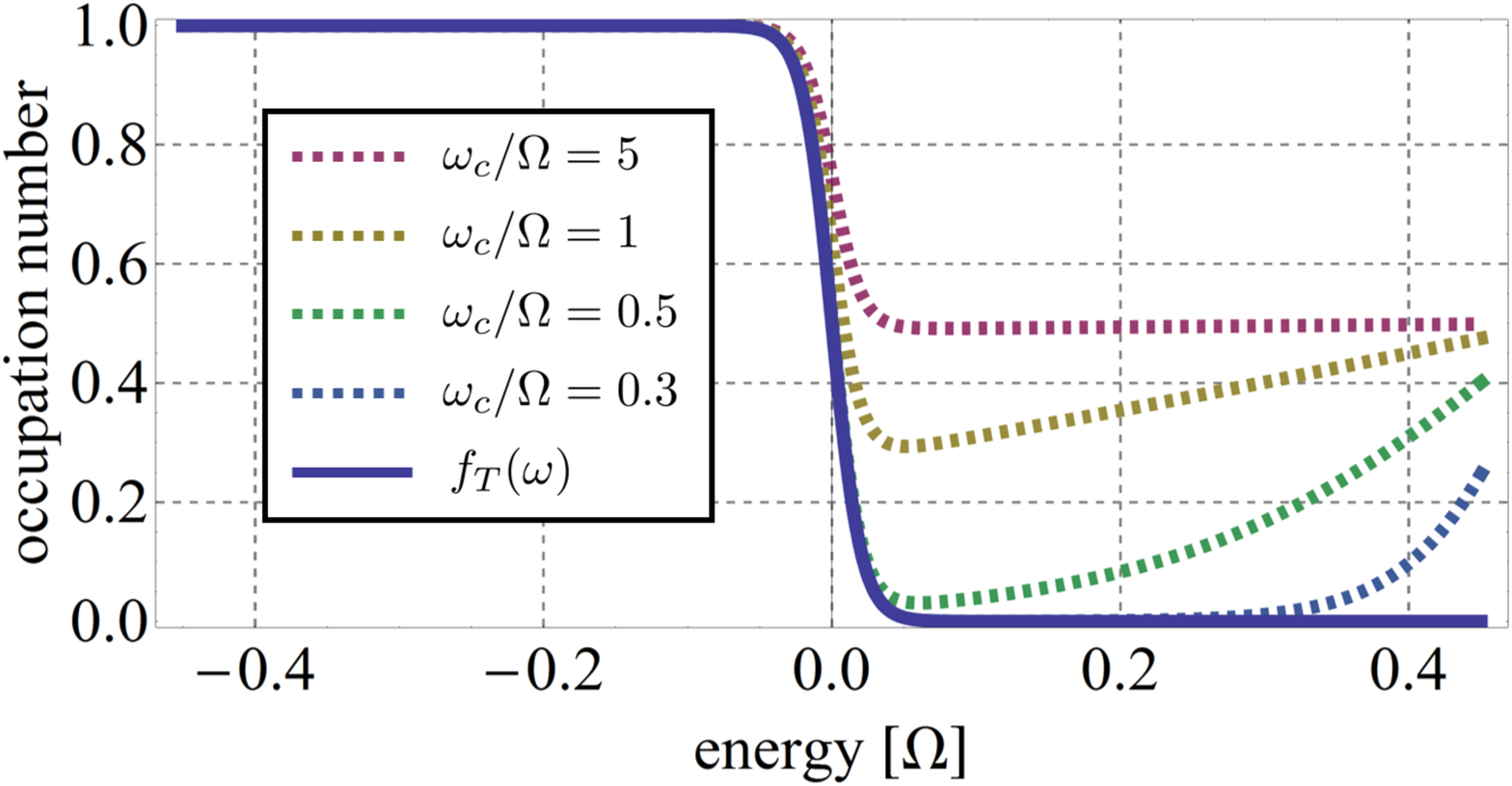}
	\caption{The occupation number of the Floquet states for the model (\ref{Eq_RabiExample}) 
			 ($\omega_{c}/\Omega=5$, red dashed line; $\omega_{c}/\Omega=1$, yellow dashed line; $\omega_{c}/\Omega=0.5$, green dashed line; $\omega_{c}/\Omega=0.3$, blue dashed).
			 The blue solid line represents the Fermi distribution function and the temperature is taken as $T/\Omega = 0.01$.}
	\label{Fig_ExampleRabi}
\end{figure}

Here, we show an example of the model we discussed above.
We consider the following fermionic toy model:
\begin{equation}
	\begin{aligned}
		&H_{\text{S}}(t)=\sum_{\alpha}H_{\alpha}(t)\\
		&\begin{aligned}
			H_{\alpha}(t)=(E_{\alpha}+\Omega)&a^{\dagger}_{\alpha,1}a_{\alpha,1}+E_{\alpha}a^{\dagger}_{\alpha,2}a_{\alpha,2}\\
							&+\Delta(e^{-i\Omega t+i\theta}a^{\dagger}_{\alpha,1}a_{\alpha,2}+h.c.)
		 \end{aligned}
	\end{aligned}\label{Eq_RabiExample}
\end{equation}
In this example, we can eliminate the time dependence of the Hamiltonian by the time-periodic unitary transformation $W(t)=\prod_{\alpha} \exp[i\Omega n_{\alpha,1} t]$.
The quasi-energy of the system is given by $\epsilon_{\alpha,\pm}=E_{\alpha}\pm\Delta$, and the contribution from the $n$th sideband can be obtained as follows:
\begin{equation}
	\Gamma_{\alpha,\pm}^{n}=\begin{cases}
								D(\epsilon_{\alpha,\pm})|\lambda_{\alpha,2}(\epsilon_{\alpha,\pm})|^{2} /4\pi & n=0\\
								D(\epsilon_{\alpha,\pm})|\lambda_{\alpha,1}(\epsilon_{\alpha,\pm})|^{2} /4\pi & n=-1\\
								0 & n\neq 0,-1.
							\end{cases}
\end{equation}
Because $W(t)$ only contains $n=0,-1$ Fourier components, there are no contributions from sidebands except for $n=0,-1$ sidebands.
We consider the case where the system-reservoir coupling is independent of the index $j$ (i.e. $\lambda_{\alpha,j}(\omega)=\lambda_{\alpha}(\omega)$) and $D(\omega)\lambda_{\alpha}(\omega)=C_{\alpha}\exp[-(\omega/\omega_{c})^2]$, where $\omega_{c}$ is the energy cutoff of the system-reservoir coupling.
The occupation number of the Floquet states $c_{\alpha,\pm}=(a_{\alpha,1}\pm e^{-i\theta}a_{\alpha,2})/\sqrt{2}$ in the asymptotic state is shown in Fig \ref{Fig_ExampleRabi}.
We can observe that the difference between the occupation number of the Floquet state and the Fermi distribution function gradually becomes small as $\omega_{c}/\Omega$ decreases.
The deviation from the Fermi distribution function is small in the range $\epsilon_{\alpha,\pm}<-T$ because $n=0,-1$ sidebands only contribute to the occupation number, that is, the occupation number of the Floquet state is written by the sum of $f_{T}(\omega)$ and $f_{T}(\omega-\Omega)$, and $f_{T}(\omega)\simeq f_{T}(\omega-\Omega)\simeq 1$ when $\omega<-T$.
As we discussed above, when the energy cutoff of the system-reservoir coupling is smaller than the frequency of the external field (see the blue dashed line), the deviation of the occupation number in the region (i) (i.e. $|\epsilon_{\alpha,\pm}|<\omega_{c}$) from the Fermi distribution function becomes small.

\section{Results for finite system-reservoir coupling}\label{Sec_FiniteCoupling}
So far we have focused on the general properties of the periodically driven open systems under infinitesimal dissipation.
Let us now investigate the finite-dissipation effect.
Especially, we consider the case where there are no energy cutoffs in the reservoirs to investigate the pure dissipation effect.
When the dispersion of the reservoir is linear $\epsilon_{k}=uk$ without cutoff ($u$ is the velocity) and the system-reservoir coupling is independent of the wave number $\lambda_{k,\alpha}=\lambda_{\alpha}$, eq. (\ref{Eq_SystemClosedEOM}) becomes Markovian \cite{Kohler2005}:
\begin{equation}
		i\frac{d}{dt}a_{\alpha}(t)=\sum_{\beta}(H_{\alpha\beta}(t)-i\Gamma_{\alpha\beta})a_{\beta}(t)+i\xi_{\alpha}(t),\label{Eq_MarkovSystemClosedEOM}
\end{equation}
where $\Gamma_{\alpha\beta}=\lambda_{\alpha}^{\ast}\lambda_{\beta}/2u$ and $\xi_{\alpha}(t)$ is the same noise term as the previous one.
We can solve this differential equation by the Floquet theorem {\it without restriction of the weak system-reservoir coupling limit}.
The Floquet theorem guarantees the existence of a set $\{\ket{\psi_{\alpha}(t)}\}$ of solutions of the time periodic Sch\"{o}dinger equation $id/dt \ket{\psi(t)}=(H(t)-i\Gamma)\ket{\psi(t)}$ which can be written as
$\ket{\psi_{\alpha}(t)}=e^{-(i\epsilon_{\alpha}+\gamma_{\alpha})t}\ket{u_{\alpha}(t)},\ \ket{u_{\alpha}(t+\tau)}=\ket{u_{\alpha}(t)}$, where $\epsilon_{\alpha}$ and $\gamma_{\alpha}$ are real and we can take $|\epsilon_{\alpha}|<\Omega/2$.
The time periodic part of the wave function $\ket{u_{\alpha}(t)}$ satisfies $(H(t)-i\Gamma-id/dt)\ket{u_{\alpha}(t)}=(\epsilon_{\alpha}-i\gamma_{\alpha})\ket{u_{\alpha}(t)}.$
Because the operator $H(t)-i\Gamma-id/dt$ is not Hermitian, we also define the left eigenstate $\bra{u^{\dagger}_{\alpha}(t)}$.
Then, the occupation number of an arbitrary basis of the system  is obtained as,
\begin{equation}
	\begin{aligned}
		\braket{a^{\dagger}_{\alpha}(t)a_{\alpha}(t)}&=\sum_{\alpha^{\prime},\beta^{\prime}}\sum_{k,k^{\prime}\in\mathbb{Z}}\braket{a_{\alpha}|u_{\alpha^{\prime}}(t)}\braket{u_{\beta^{\prime}}(t)|a_{\alpha}}e^{i(k-k^{\prime})\Omega t}\\
											&\times \int \frac{d\omega}{\pi}\frac{\braket{u^{\dagger(k)}_{\alpha^{\prime}}|\Gamma|u^{\dagger(k^{\prime})}_{\beta^{\prime}}}f_{T}(\omega)}{(\omega-\epsilon^{(k)}_{\alpha^{\prime}}+i\gamma_{\alpha^{\prime}})(\omega-\epsilon^{(k^{\prime})}_{\beta^{\prime}}-i\gamma_{\beta^{\prime}})},
	\end{aligned}\label{Eq_MarkovOccupationNum}
\end{equation}
where $\ket{u_{\alpha}^{(n)}},\ \ket{u^{\dagger(n)}_{\alpha}}$ are the $n$th Fourier coefficients of $\ket{u_{\alpha}(t)},\ \ket{u^{\dagger}_{\alpha}(t)}$ and $\epsilon_{\alpha}^{(n)}=\epsilon_{\alpha}+n\Omega$.
The formula (\ref{Eq_MarkovOccupationNum}) is exact and applicable for any $\Omega$ and $\Gamma$.
If we take the weak system-reservoir coupling limit $\gamma_{\alpha}\rightarrow 0$, eq. (\ref{Eq_MarkovOccupationNum}) is reduced to eq. (\ref{Eq_FlouetOccupationNum}).
Therefore, the expression (\ref{Eq_MarkovOccupationNum}) is an extension of eq. (\ref{Eq_FlouetOccupationNum}) to the case with finite-dissipation effect.
When $H(t)$ is time independent, $\ket{u_{\alpha}^{(n)}}=\ket{u_{\alpha}^{\dagger(n)}}=0$ for $n{\neq}0$.
Therefore the formula (\ref{Eq_MarkovOccupationNum}) can be interpreted as that for an equilibrium system which has an infinite number of energy bands $\{ \ket{u^{(n)}_{\alpha}},\ \epsilon^{(n)}_{\alpha} \}$ except for the phase factor $e^{-ikt}$.
These energies are broadened by the width $\gamma_{\alpha}$, but the states in the different sidebands are not orthogonal even though the system-reservoir coupling is taken as zero.
Namely, different sidebands are not independent but just duplicated \cite{Tien1963}.

To simplify the analysis, we consider the high frequency regime of the external driving.
We take the initial temperature as zero.
Both the system-reservoir coupling and the energy difference between the different sidebands is much smaller than the frequency of the external driving.
The corresponding conditions read:
\begin{equation}
	\frac{\gamma_{\alpha}}{\Omega},\ \frac{\gamma_{\alpha}}{\min_{\alpha,\beta}|\Omega-(\epsilon_{\alpha}-\epsilon_{\beta})|}\rightarrow 0.\label{Eq_HighFrequencyCondition}
\end{equation}
If the first condition is satisfied, the second is also satisfied when the Magnus expansion applicable.
The second condition means that the overlap between the different sidebands which is caused by the width $\gamma_{\alpha}$ vanishes.
In this limit, the occupation number can be obtained as,
\begin{equation}
	\begin{aligned}
		\braket{a^{\dagger}_{\alpha}(t)a_{\alpha}(t)}&=\sum_{\alpha^{\prime},\beta^{\prime}}\braket{a_{\alpha}|u_{\alpha^{\prime}}(t)}\braket{u_{\beta^{\prime}}(t)|a_{\alpha}}\\
											&\times \int \frac{d\omega}{\pi} \frac{\Gamma_{\alpha^{\prime}\beta^{\prime}}^{(0)}f_{T=0}(\omega)+\Gamma_{\alpha^{\prime}\beta^{\prime}}^{(-)}}{(\omega-\epsilon_{\alpha^{\prime}}+i\gamma_{\alpha^{\prime}})(\omega-\epsilon_{\beta^{\prime}}-i\gamma_{\beta^{\prime}})}
	\end{aligned}
\label{Eq_MarkovHighFrequencyRegime}
\end{equation}
where $\Gamma_{\alpha\beta}^{(k)}=\braket{u^{\dagger(k)}_{\alpha}|\Gamma|u^{\dagger(k)}_{\beta}}$ and $\Gamma^{(-)}_{\alpha\beta}=\sum_{k<0}\Gamma^{(k)}_{\alpha\beta}$.
We can consider that the second term $\Gamma^{(-)}_{\alpha\beta}$ in eq. (\ref{Eq_MarkovHighFrequencyRegime}) represents the excitation from the $T=0$ state because there is only $0$th sideband when $H(t)$ is time independent.
When we use the Magnus expansion to obtain the quasienergy spectrum of the system, $\Gamma^{(-)}_{\alpha\beta}=O((A/\Omega)^2)$, where $A$ is the amplitude of the external driving \cite{Iadecola2015b}.
This is of the same order of magnitude as the contribution from the sidebands to the occupation number in the weak system-reservoir coupling limit, thereby implying that we cannot suppress excitations by simply considering a stronger dissipation.
This is because the high energy excitations can be created through the reservoir by the external driving.

\section{Summary}\label{Sec_Summary}
To summarize, we have investigated fundamental properties of periodically driven open quantum systems, which are independent of the detailed nature of the reservoir, and the finite dissipation effect particularly focusing on free fermionic systems.
We have revealed the following two properties.
(i) Low-energy properties are equivalent to those of the Floquet Gibbs state when the frequency of the external driving is much larger than the energy cutoff of the system-reservoir coupling even though there is no fine tuning of the system-reservoir coupling.
(ii) Finite system-reservoir coupling cannot suppress the excitations in the system because external drivings can create the excitations through the reservoir if the frequency of the external driving is smaller than the energy cutoff of the system-reservoir coupling.
Therefore, the external driving whose frequency is larger than the energy cutoff of the system-reservoir coupling is required to obtain the thermodynamic properties of a effective Hamiltonian.

As further studies, clarifying the general feature of asymptotic states of the periodically driven open many body quantum systems and non-Markovian quantum systems is an intriguing and important problem as we have investigated Markovian free fermionic systems in both infinitesimal and finite system-reservoir coupling cases.

\begin{acknowledgments}
This work was partly supported by a Grand-in-Aid for Scientific Research on Innovative Areas (JSPS KAKENHI Grant No. JP15H05855) and also JSPS KAKENHI (No. JP16K05501).
\end{acknowledgments}

\appendix
\begin{widetext}

\section{Derivation of eq. (\ref{Eq_FlouetOccupationNum})}\label{appendix_a}
We here derive the occupation number and the time correlation functions of the Floquet state in the weak system-reservoir coupling limit.
The solution of eq. (\ref{Eq_FloquetEOM}) is eq. (\ref{Eq_SolutionOfEOM}):
\begin{gather}
	c_{\alpha}(t)=\sum_{k,\beta}\sum_{n\in\mathbb{Z}}\frac{1}{\sqrt{L}}\frac{U^{(n)\dagger}_{\alpha\beta}(\lambda_{k,\beta})^{\ast}e^{-i(\epsilon_{k}-n\Omega)t}}{\epsilon_{k}-(\epsilon_{\alpha}+n\Omega)+i\Gamma_{\alpha}^{\prime}(n\Omega-\epsilon_{k})}A_{k} \label{App_SolutionOfEOM} \\
	\Gamma_{\alpha}^{\prime}(\omega)=\sum_{n\in\mathbb{Z}}\int_{0}^{\infty}dt\Gamma_{\alpha\alpha}^{nn}(t)e^{i(\omega-n\Omega)t}
\end{gather}
Then, the time correlation function of the Floquet state is written in the following way:
\begin{equation}
	\braket{c^{\dagger}_{\alpha}(t)c_{\beta}(t^{\prime})}=\sum_{n,m\in\mathbb{Z}}\int_{-\infty}^{\infty}d\omega \frac{f_{T}(\omega)\tilde{\Gamma}^{mn}_{\beta\alpha}(\omega)e^{i\omega(t-t^{\prime})+im\Omega t^{\prime}-in\Omega t}}{(\omega-(\epsilon_{\beta}+m\Omega)+i\Gamma_{\beta}^{\prime}(m\Omega-\omega))(\omega-(\epsilon_{\alpha}+n\Omega)-i\Gamma_{\alpha}^{\prime \ast}(n\Omega-\omega))} \label{App_RawTimeCorrelationFunc}
\end{equation}
Only the $n=m$ and $\alpha=\beta$ terms contribute to eq. (\ref{App_RawTimeCorrelationFunc}) in the weak system-reservoir coupling limit ($\lambda \rightarrow 0$) because we assumed the nondegeneracy and nonresonance ($\epsilon_{\alpha}-\epsilon_{\beta}+n\Omega=0 \Leftrightarrow \alpha=\beta,\ n=0$) in the derivation of eq. (\ref{App_SolutionOfEOM}).
Therefore, we obtain the following expression:
\begin{equation}
	\begin{aligned}
	\braket{c^{\dagger}_{\alpha}(t)c_{\beta}(t^{\prime})}&=\sum_{n\in\mathbb{Z}}\int_{-\infty}^{\infty}d\omega \frac{f_{T}(\omega)\tilde{\Gamma}^{nn}_{\alpha\alpha}(\omega)e^{i(\omega-n\Omega)(t-t^{\prime})}}{|\omega-(\epsilon_{\alpha}+n\Omega)+i\Gamma_{\alpha}^{\prime}(n\Omega-\omega)|^{2}}\delta_{\alpha,\beta}\\
														&=\frac{\sum_{n\in\mathbb{Z}}\pi\tilde{\Gamma}^{nn}_{\alpha\alpha}(\epsilon_{\alpha}+n\Omega)f_{T}(\epsilon_{\alpha}+n\Omega)e^{i\epsilon_{\alpha}(t-t^{\prime})}}{\text{Re}[\Gamma_{\alpha}^{\prime}(-\epsilon_{\alpha})]}\delta_{\alpha,\beta}
	\end{aligned}
\end{equation}
We here applied the relation
\begin{equation}
	\lim_{\delta\downarrow 0}\int_{-\infty}^{\infty}\frac{d\omega}{\pi}\frac{\delta}{\omega^{2}+\delta^{2}}F(\omega)=\int_{-\infty}^{\infty}d\omega\delta(\omega)F(\omega)=F(0).
\end{equation}
By utilizing the property $\Gamma_{\alpha\alpha}^{nn}(\omega)>0$, we can prove the relation $\text{Re}[\Gamma_{\alpha}^{\prime}(-\epsilon_{\alpha})]=\pi\sum_{n\in\mathbb{Z}}\tilde{\Gamma}_{\alpha\alpha}^{nn}(\epsilon_{\alpha}+n\Omega)$, then we obtain the following expression:
\begin{equation}
	\braket{c^{\dagger}_{\alpha}(t)c_{\beta}(t^{\prime})}=\frac{\sum_{n\in\mathbb{Z}}\tilde{\Gamma}^{nn}_{\alpha\alpha}(\epsilon_{\alpha}+n\Omega)f_{T}(\epsilon_{\alpha}+n\Omega)}{\sum_{n\in\mathbb{Z}}\tilde{\Gamma}^{nn}_{\alpha\alpha}(\epsilon_{\alpha}+n\Omega)}e^{i\epsilon_{\alpha}(t-t^{\prime})}\delta_{\alpha,\beta} \label{App_TimeCorrelationFunc}
\end{equation}
We can obtain the other time correlation functions in the same way and they have the same structure as eq. (\ref{App_TimeCorrelationFunc}).
\end{widetext}

\bibliography{manuscript}

\end{document}